# Conductivity low-energy asymptotics for monolayer graphene

Natalie E. Firsova

*199178,   St. Petersburg, Russia*

**Abstract.** Electron scattering problem in the monolayer graphene with short-range impurities is considered. The main novel element in the suggested model is the band asymmetry of the defect potential in the 2+1-dimensional Dirac equation. This asymmetry appears naturally if the defect violates the symmetry between sublattices. Our goal in the present paper is to take into account a local band asymmetry violation arising due to the defect presence. We analyze the effect of the electron scattering on the electronic transport parameters in the monolayer graphene. The explicit exact formulae obtained for $S$-matrix for the suggested $\delta-$shell potential model allowed us to study the asymptotic behavior of such characteristics as scattering phases, transport cross section, the transport relaxation time and the conductivity for small values of the Fermi energy. The obtained results are in a good agreement with the experimental data which shows that the considered model is reasonable.

**Introduction**

During the last years much attention was paid to the problem of the electronic spectrum of graphene (see the review [1]). Its 2D-structure and the presence of the cone points in the electronic spectrum make actual a comprehensive study of the external field effect on the spectrum and other characteristics of the electronic states described by the Dirac equation in the 2+1 space-time. We consider in the present paper the transport phenomena in the 2+1 Dirac equation model of the monolayer graphene due to the short-range perturbation. We do not take into account the inter-valley transitions. Particular attention to this case stems from the effectiveness of short-range scatterers in contrast to the long-range ones: an effect of the latter is suppressed by the Klein paradox [2]. Short-range potential impurities in graphene were considered in papers [3-6]. In [6], for instance, electrons were assumed to be confined in a quantum dot where the dot was represented by a $\delta^{(2)}(\vec{r})$-potential well. Artificially representing the quantum dot by such a strongly singular potential leads to divergences in the Lippmann-Schwinger equation. To overcome the problem in [7] the $\delta$-shell model was suggested which removed the singularity. The Dirac equation for the $\delta$-shell potential is free of such divergences. In [7] a new model of the short-range impurities in graphene was considered where the shell delta function potential form was suggested taking into account for the first time the obvious fact that the Kohn-Luttinger matrix elements of the short-range perturbations calculated on the upper and lower band wave functions are not equal in a general case. This means that the perturbation must be generically described by a Hermitian matrix. In [7] the diagonal matrix case corresponding to a presence of the chemical potential and the mass perturbation was studied taking firstly into account a local band symmetry violation arising due to the defect presence. In [7] for the model the characteristic equation for eigenvalues and resonances was obtained describing their dependence on the perturbation parameters. In [8] in the framework of the model suggested in [7] the electron scattering was studied and the exact analytical formula for $S$-matrix was found.

In the present paper we analyze the effect of the electron scattering studied in [8] on the electronic transport in the monolayer graphene. We compare our theoretical results with the available experimental data.

**The main results**

The Dirac equation describing electronic states in monolayer graphene in the framework of the model described above (see [7-8]) reads:

$$\left[-i\hbar v_F \sum_{\mu=1}^{2} \gamma_\mu \partial_\mu - \gamma_0 (m+\delta m) v_F^2\right]\psi = (E-V)\psi, \quad (1)$$

where $v_F$ is the Fermi velocity of the band electron, $\gamma_\mu$ are the Dirac matrices,

$$\gamma_0 = \sigma_3, \quad \gamma_1 = \sigma_1, \quad \gamma_2 = i\sigma_2,$$

$\sigma_j$ are the Pauli matrices, $2mv_F^2 = E_g$ is the electronic bandgap, and $\psi(\vec{r})$ is the two-component spinor. The spinor structure takes into account the two-sublattice structure of the graphene.

We consider firstly the gapless case $m = 0$. Then we have

$$\left[-i\hbar v_F \sum_{\mu=1}^{2} \gamma_\mu \partial_\mu - \gamma_0 \delta m v_F^2\right]\psi = (E-V)\psi. \quad (2)$$

To treat this equation mathematically we make (2) dimensionless dividing it by. $\hbar v_F k_F$ We obtain:

$$\left[-i\sum_{\mu=1}^{2} \gamma_\mu \partial_\mu - \gamma_0 \delta \tilde{m}\right]\psi = (\tilde{E} - \tilde{V})\psi. \quad (3)$$

Here $\delta\tilde{m} = \delta m v_F (\hbar k_F)^{-1}$, $\tilde{E} = E(\hbar v_F k_F)^{-1}$, $\tilde{V} = V(\hbar v_F k_F)^{-1}$. $\tilde{V}(\vec{r})$ is a local perturbation of the chemical potential. A local mass (gap) perturbation $\delta\tilde{m}$ related to a local sublattices symmetry violation which can be induced for instance by defects in the graphene film or in the substrate (see [9]). We consider here the shell delta function model of the perturbation

$$\delta\tilde{m}(\tilde{r}) = -b\delta(\tilde{r} - \tilde{r}_0), \quad \tilde{V}(\tilde{r}) = -a\delta(\tilde{r} - \tilde{r}_0) \quad (4)$$

where $\tilde{r} = rk_F$ and $\tilde{r}_0 = r_0 k_F$ are respectively the polar coordinate radius and the perturbation radius. The finite radius $r_0$ plays a role of the regulator and it is necessary in order to exclude deep states of the atomic energy scale. The finite perturbation radius $r_0$ leads to the quasi-momentum space form-factor proportional to the Bessel function that justifies our neglect of transitions between the Brillouin band points $K$ and $K'$, [10]. The perturbation matrix elements

$$diag(\tilde{V}_1, \tilde{V}_2)\, \delta(\tilde{r} - \tilde{r}_0)$$

are related to the $a, b$ parameters as follows

$$-\tilde{V}_1 = a+b, \qquad -\tilde{V}_2 = a-b$$

Solving the Dirac equation (3) in the regions $0 \leq \tilde{r} < \tilde{r}_0$, $\tilde{r}_0 < \tilde{r} < \infty$ and matching these solutions at the circumference of the circle of the radius $\tilde{r} = \tilde{r}_0$ the characteristic equation was obtained for eigenvalues and resonances, [7]. Calculating the ratio of the outgoing and ingoing waves the formulae for $S$-matrix components were found in the angular momentum representation, [8]:

$$S_j(k\tilde{r}_0) = -\frac{F_j^{(2)}(k\tilde{r}_0)}{F_j^{(1)}(k\tilde{r}_0)}, \qquad j = \pm\frac{1}{2}, \pm\frac{3}{2}, ..., \qquad k = E \qquad (5)$$

where $F_j^{(\alpha)}(k\tilde{r}_0)$, $\alpha=1,2$, is given by the formula

$$F_j^{(\alpha)}(k\tilde{r}_0) = \left[ I_{j-\frac{1}{2}}(k\tilde{r}_0) H_{j+\frac{1}{2}}^{(\alpha)}(k\tilde{r}_0) - I_{j+\frac{1}{2}}(k\tilde{r}_0) H_{j-\frac{1}{2}}^{(\alpha)}(k\tilde{r}_0) \right] -$$

$$T(a,b)\left[ (a-b) I_{j+\frac{1}{2}}(k\tilde{r}_0) H_{j+\frac{1}{2}}^{(\alpha)}(k\tilde{r}_0) + (a+b) I_{j-\frac{1}{2}}(k\tilde{r}_0) H_{j-\frac{1}{2}}^{(\alpha)}(k\tilde{r}_0) \right],$$

$$\alpha = 1, 2, \qquad j = \pm\frac{1}{2}, \pm\frac{3}{2}, ... \qquad (6)$$

Here

$$T(a,b) = \begin{cases} tg\sqrt{a^2-b^2} / \sqrt{a^2-b^2} & \text{if } a^2 > b^2 \\ th\sqrt{b^2-a^2} / \sqrt{b^2-a^2} & \text{if } b^2 > a^2 \end{cases} \qquad (7)$$

where we choose the principal value of the roots. Since for Hankel functions we have for real variables $\overline{H_n^{(2)}(x)} = H_n^{(1)}(x)$ the scattering matrix (5), (6) is unitary on the continuum spectrum $\text{Im}\,\tilde{E} = 0$. Using the relation [11]

$$\left[ I_{j-\frac{1}{2}}(k\tilde{r}_0) N_{j+\frac{1}{2}}(k\tilde{r}_0) - I_{j+\frac{1}{2}}(k\tilde{r}_0) N_{j-\frac{1}{2}}(k\tilde{r}_0) \right] = -\frac{2}{\pi k\tilde{r}_0}, \qquad (8)$$

we can rewrite (6) in the form

$$F_j^{(\alpha)}(k\tilde{r}_0) = (-1)^\alpha \frac{2i}{\pi k\tilde{r}_0} - T(a,b)\left[ (a-b) I_{j+\frac{1}{2}}(k\tilde{r}_0) H_{j+\frac{1}{2}}^{(\alpha)}(k\tilde{r}_0) + (a+b) I_{j-\frac{1}{2}}(k\tilde{r}_0) H_{j-\frac{1}{2}}^{(\alpha)}(k\tilde{r}_0) \right],$$

$$\alpha = 1, 2, \qquad j = \pm\frac{1}{2}, \pm\frac{3}{2}, ... \qquad (9)$$

So the poles of the scattering matrix (5) i.e. eigenvalues and resonances are determined as solutions of the characteristic equation

$$F_j^{(1)}(k\tilde{r}_0) = 0, \qquad j = \pm\frac{1}{2}, \pm\frac{3}{2}, \ldots \qquad (10)$$

or

$$T(a,b)\left[(a-b)I_{j+\frac{1}{2}}(k\tilde{r}_0)H_{j+\frac{1}{2}}^{(\alpha)}(k\tilde{r}_0) + (a+b)I_{j-\frac{1}{2}}(k\tilde{r}_0)H_{j-\frac{1}{2}}^{(\alpha)}(k\tilde{r}_0)\right] = -\frac{2i}{\pi k\tilde{r}_0},$$

$$j = \pm\frac{1}{2}, \pm\frac{3}{2} \ldots \qquad (11)$$

Using the relations $H_n^{(1)}(x) = I_n(x) + iN_n(x)$, $H_n^{(1)}(x) = I_n(x) - iN_n(x)$ we can write S-matrix (5) as follows

$$S_j(k\tilde{r}_0) = -\frac{A_j(k\tilde{r}_0) - iB_j(k\tilde{r}_0)}{A_j(k\tilde{r}_0) + iB_j(k\tilde{r}_0)} = \frac{B_j(k\tilde{r}_0) + iA_j(k\tilde{r}_0)}{B_j(k\tilde{r}_0) - iA_j(k\tilde{r}_0)}, \qquad j = \pm\frac{1}{2}, \pm\frac{3}{2}, \ldots \qquad (12)$$

Therefore it can be presented in the standard form

$$S_j(k\tilde{r}_0) = \exp(2i\delta_j(k\tilde{r}_0)), \qquad j = \pm\frac{1}{2}, \pm\frac{3}{2}, \ldots \qquad (13)$$

where for the scattering phases we have

$$\delta_j(k\tilde{r}_0) = \operatorname{arctg}\frac{A_j(k\tilde{r}_0)}{B_j(k\tilde{r}_0)}, \qquad j = \pm\frac{1}{2}, \pm\frac{3}{2}, \ldots \qquad (14)$$

The functions $A_j(k\tilde{r}_0), B_j(k\tilde{r}_0)$, $j = \pm\frac{1}{2}, \frac{3}{2}, \ldots$ are determined (see (9)) as follows

$$A_j(k\tilde{r}_0) = -T(a,b)\left[(a-b)I_{j+\frac{1}{2}}^2(k\tilde{r}_0) + (a+b)I_{j-\frac{1}{2}}^2(k\tilde{r}_0)\right] \qquad (15)$$

$$B_j(k\tilde{r}_0) = -\frac{2}{\pi k\tilde{r}_0} - T(a,b)\left[(a-b)I_{j+\frac{1}{2}}(k\tilde{r}_0)N_{j+\frac{1}{2}}(k\tilde{r}_0) + (a+b)I_{j-\frac{1}{2}}(k\tilde{r}_0)N_{j-\frac{1}{2}}(k\tilde{r}_0)\right],$$

$$j = \pm\frac{1}{2}, \pm\frac{3}{2}, \ldots \qquad (16)$$

**Lemma1.** Scattering phases $\delta_j(k\tilde{r}_0)$, $j = \pm\frac{1}{2}, \pm\frac{3}{2}, \ldots$ of the $S_j(k\tilde{r}_0)$-matrix of the Dirac equation (3) near the Dirac point $k = 0$ that is for low energies (small momentum) i.e. for $k\tilde{r}_0 \to 0$ have asymptotics uniformly on a set of $j$

$$\delta_{\pm j}(k\tilde{r}_0) = \frac{\pi T(a,b)(a \pm b)}{\left[\left(j - \frac{1}{2}\right)!\right]^2}\left(\frac{k\tilde{r}_0}{2}\right)^{2j}[1 + o(1)], \qquad k\tilde{r}_0 \to 0, \qquad j = \frac{1}{2}, \frac{3}{2}, \ldots, \qquad (17)$$

**Proof.** From (14)-(16) we see

$$tg\ \delta_j(k\tilde{r}_0) = \frac{\pi k\tilde{r}_0}{2}T(a,b)\left[(a-b)I^2_{j+\frac{1}{2}}(k\tilde{r}_0)+(a+b)I^2_{j-\frac{1}{2}}(k\tilde{r}_0)\right]\times$$

$$\times\left[1+\frac{\pi k\tilde{r}_0}{2}T(a,b)\left[(a-b)I_{j+\frac{1}{2}}(k\tilde{r}_0)N_{j+\frac{1}{2}}(k\tilde{r}_0)+(a+b)I_{j-\frac{1}{2}}(k\tilde{r}_0)N_{j-\frac{1}{2}}(k\tilde{r}_0)\right]\right]^{-1} \quad (18)$$

$$j=\pm\frac{1}{2},\pm\frac{3}{2},\ldots$$

So the asymptotic behavior of the scattering phases $\delta_j(k\tilde{r}_0)$ at $k\tilde{r}_0 \to 0$ can be obtained expanding the cylinder functions for small arguments (see [11])

$$I_n(x) \sim (1/n!)(x/2)^n, \qquad n=0,1,2\ldots, \quad (19)$$

$$N_n(x) \sim \begin{cases} -(\Gamma(n)/\pi)(2/x)^n & \text{for } n>0 \\ (2/\pi)\log(\gamma_E x/2) & \text{for } n=0 \end{cases} \quad (20)$$

where $\gamma_E \approx 0{,}577$ is the Eyler–Mascerone constant and $\Gamma(n)$ is the *gamma*-function. From (18)-(20) we obtain asymptotic uniform on the set of $j$

$$tg\left(\delta_{\pm j}(k\tilde{r}_0)\right) = \frac{\pi T(a,b)(a\pm b)}{\left[\left(j-\frac{1}{2}\right)!\right]^2}\left(\frac{k\tilde{r}_0}{2}\right)^{2j}[1+o(1)], \quad k\tilde{r}_0 \to 0, \quad j=\frac{1}{2},\frac{3}{2},\ldots, \quad (21)$$

Expanding the function $arctg\left(tg\delta_{\pm j}(k\tilde{r}_0)\right)$ $k\tilde{r}_0 \to 0$ we find

$$\delta_{\pm j}(k\tilde{r}_0) = \frac{\pi T(a,b)(a\pm b)}{\left[\left(j-\frac{1}{2}\right)!\right]^2}\left(\frac{k\tilde{r}_0}{2}\right)^{2j}[1+o(1)]\left[1+0\left(\left(\frac{k\tilde{r}_0}{2}\right)^{4j}\right)\right], \quad k\tilde{r}_0 \to 0, \quad j=\frac{1}{2},\frac{3}{2},\ldots,$$

Hence we come to (17). ∎

Let us now define $\delta_j(k_F r_0) = \delta_j(k\tilde{r}_0)$, $S_j(k_F r_0) = S_j(k\tilde{r}_0)$, $j=\pm\frac{1}{2},\pm\frac{3}{2}\ldots$ Then the transport cross section can be written in terms of the scattering phases

$$\Sigma_{tr}(k_F, r_0) = \frac{2}{k_F}(\Sigma_1)_{tr}(k_F r_0) \quad (22)$$

where

$$(\Sigma_1)_{tr}(k_F r_0) = (\Sigma_1)_{tr}(k\tilde{r}_0) = \sum_{j=\pm\frac{1}{2},\pm\frac{3}{2},\ldots} \left[\sin\left(\delta_{j+1}(k\tilde{r}_0) - \delta_j(k\tilde{r}_0)\right)\right]^2. \tag{23}$$

The transport relaxation time $\tau_{tr}$ can be calculated using the following relation

$$1/\tau_{tr}(k_F r_0) = N_i v_F \Sigma_{tr}(k_F, r_0) \tag{24}$$

The Boltzmannian conductivity is determined by the formula

$$\sigma(E_F, r_0) = \frac{e^2}{h}\left(E_F \tau_{tr}(k_F, r_0)/\hbar\right) \tag{25}$$

**Theorem1** We have for the transport cross section of the problem (2) we have

$$\Sigma_{tr}(k_F, r_0) = \pi^2 T(a,b)^2 \left(a^2 + 3b^2\right) k_F r_0^2 \left[1 + o(1)\right], \quad k_F r_0 \ll 1 \tag{26}$$

The conductivity for the problem (2) for low energies has asymptotics

$$\sigma = \sigma_0 \left[1 + o(1)\right], \quad k_F r_0 \ll 1, \tag{27}$$

where

$$\sigma_0 = \frac{e^2}{h}\left[\pi^2 T(a,b)^2 \left(a^2 + 3b^2\right) N_i r_0^2\right]^{-1} \tag{28}$$

**Proof.** From Lemma1 we see that $\delta_{\pm j}(k\tilde{r}_0) \to 0$, when $k\tilde{r}_0 \to 0$, $j = \frac{1}{2}, \frac{3}{2}, \ldots$ It means that expanding we obtain

$$\sin\left(\delta_{j+1}(k\tilde{r}_0) - \delta_j(k\tilde{r}_0)\right) = \left(\delta_{j+1}(k\tilde{r}_0) - \delta_j(k\tilde{r}_0)\right)\left[1 + O\left(\left(\delta_{j+1}(k\tilde{r}_0) - \delta_j(k\tilde{r}_0)\right)^2\right)\right], \quad k\tilde{r}_0 \to 0,$$

$$j = \pm\frac{1}{2}, \pm\frac{3}{2}, \ldots$$

or

$$\sin\left(\delta_{j+1}(k\tilde{r}_0) - \delta_j(k\tilde{r}_0)\right) = \left(\delta_{j+1}(k\tilde{r}_0) - \delta_j(k\tilde{r}_0)\right)\left[1 + o(1)\right], \quad k\tilde{r}_0 \to 0. \quad j = \pm\frac{1}{2}, \pm\frac{3}{2}, \ldots$$

So we can find that (see (23))

$$(\Sigma_1)_{tr}(k\tilde{r}_0) = (\Sigma_2)_{tr}(k\tilde{r}_0)\left[1 + o(1)\right], \quad k\tilde{r}_0 \to 0. \tag{29}$$

where

$$(\Sigma_2)_{tr}(k\tilde{r}_0) = \sum_{j=\pm\frac{1}{2},\pm\frac{3}{2},\ldots} \left(\delta_{j+1}(k\tilde{r}_0) - \delta_j(k\tilde{r}_0)\right)^2 \tag{30}$$

Transforming (30) we get

$$(\Sigma_2)_{tr}(k\tilde{r}_0) = 2\left\{\delta_{\frac{1}{2}}^2 + \delta_{-\frac{1}{2}}^2 - \delta_{\frac{1}{2}}\delta_{-\frac{1}{2}} - \sum_{j=\frac{1}{2},\frac{3}{2}\ldots}\left[\left(\delta_j\delta_{j+1} + \delta_{-j}\delta_{-j-1}\right) - \left(\delta_{j+1}^2 + \delta_{-j-1}^2\right)\right]\right\}$$

Using asymptotics from lemma1 we find

$$(\Sigma_2)_{tr} = 2\pi^2 T(a,b)^2(a^2+3b^2)\left(\frac{k\tilde{r}_0}{2}\right)^2[1+o(1)], \qquad k\tilde{r}_0 \to 0.$$

Substituting this equation into (29) we have

$$(\Sigma_1)_{tr}(k\tilde{r}_0) = \pi^2 T(a,b)^2(a^2+3b^2)\frac{(k\tilde{r}_0)^2}{2}[1+o(1)], \qquad k\tilde{r}_0 \to 0. \tag{31}$$

Hence using (22), (23) we obtain (26).

From (24)-(26) we see

$$E_F\tau_{tr}/\hbar = \left[\pi^2 T(a,b)^2(a^2+3b^2)N_i r_0^2\right]^{-1}[1+o(1)], \qquad k_F r_0 \ll 1, \tag{32}$$

And consequently (see (25))

$$\sigma = \frac{e^2}{h}\left[\pi^2 T(a,b)^2(a^2+3b^2)N_i r_0^2\right]^{-1}[1+o(1)], \qquad k_F r_0 \ll 1, \tag{33}$$

So the theorem1 is proved. ∎

**Discussion**

The results found above were obtained in the framework of the model assuming the shell delta form of the potential (2)-(4). For this assumption to be reasonable the perturbation radius $r_0$ should be much less than the wavelength $2\pi/k_F$ i.e. there should be satisfied the estimate

$$k_F r_0 \ll 1 \tag{34}$$

We see that this physical condition (34) of the correctness of the considered $\delta$-shell model guarantees also the correctness of the obtained asymptotics (26), (27). From the point of view of graphene physics the assumption (34) means that the circle of the radius contains no more than one unit cell of the graphene hexagon lattice. So the principal term in our asymptotics (26), (27) describes the physics well enough. For instance the different pairs $(a, b)$ of intensities satisfying the relation

$$T(a,b)^2(a^2+3b^2) = C$$

with the same constant produce the same conductivity.

Consider now the mobility which can be defined as the ratio

$$\mu = \sigma/(en), \qquad (35)$$

where the carrier density at low temperature is determined as follows

$$n = N/S = \frac{1}{2\pi}\left(\frac{E_F}{\hbar v_F}\right)^2 \qquad (36)$$

Substituting (33), (36) into (35) we find

$$\mu = \frac{\mu_0}{E_F^2}, \qquad \mu_0 = e\hbar v_F^2 \left[\pi^2 T(a,b)^2 (a^2+3b^2) N_i r_0^2\right]^{-1}[1+o(1)], \qquad k_F r_0 \ll 1, \qquad (37)$$

Notice that the obtained asymptotics for the mobility is in a good agreement with the experimental results published by Bolotin et al in [12].

Consider now the case $m \neq 0$. Following the same procedure that we used for the case $m=0$ we could see finally that the form of asymptotics for mobility greatly differs from the one we obtained for the case $m=0$ and from the corresponding experimental results. Thus it is clear that in the studied in [12] suspended monolayer graphene samples there is $m=0$ i.e., there is no gap in the spectrum.

**Conclusion** In the present paper in the framework of the model suggested in [7-8] and using the exact analytic formula for S-matrix found in [8] we obtained the asymptotics near the Dirac point for scattering phases, transport cross-section, conductivity and mobility. We found out that these asymptotics are in a good agreement with the experimental results in the case $m=0$, while in the case $m \neq 0$ the theoretical asymptotics are very far from the experimental data. This contradiction means that as a matter of fact in the samples studied experimentally in [12] there is $m=0$. So the experimental study of the sample compared to our theoretical results indicates whether there is a gap in spectrum or there is not.